\documentclass[aps,pra,twocolumn,float]{revtex4}
\usepackage{graphicx}
\usepackage{bm}
\begin{document}

\title{Coherent Dark  Resonances in Atomic Barium}

\author{U. Dammalapati, S. De, K. Jungmann and L. Willmann}
\address{Kernfysisch Versneller Instituut, University of Groningen,
                 Zernikelaan 25, 9747 AA Groningen, The Netherlands}

\begin{abstract}
The observation of dark-resonances in the two-electron atom barium
and their influence on optical cooling is reported. In heavy
alkali earth atoms, i.e. barium or radium, optical cooling can be
achieved using n$^1$S$_0$-n$^1$P$_1$ transitions and optical
repumping from the low lying n$^1$D$_2$ and n$^3$D$_{1,2}$ states
to which the atoms decay with a high branching ratio. The cooling
and repumping transition have a common upper state. This leads to
dark resonances and hence make optical cooling less inefficient.
The experimental observations can be accurately modelled by the
optical Bloch equations. Comparison with experimental results
allows us to extract relevant parameters for effective laser
cooling of barium.
\end{abstract}

\maketitle


The pioneering work on optical cooling and trapping of neutral
atoms as well as important advances made since then have mainly
been carried out with atomic systems which have a strong optical
transition connected to the ground or metastable state. Although a
two-level system would be ideal, even optically rather simple
accessible atoms, such as stable alkali isotopes, have hyperfine
structure. This requires (re-)pumping of the atomic population
into the state of highest angular momentum in order to realize an
effective two-level system.

Some atoms with complex atomic structure have a unique potential
for precision tests of fundamental symmetries
\cite{Akeson2006,Jungmann2005}. This can be exploited best with
cooled and trapped atoms. Among these systems is the radium (Ra)
atom, which offers a strong enhancement of a potential permanent
electron Electric Dipole Moment (EDM) due to the unique close
proximity of the 7s6d$^3$P$_1$ and 7s6d$^3$D$_2$ levels
\cite{Dzuba2001}. In addition it provides an excellent opportunity
to find a potential nucleon EDM in some isotopes where the nuclei
are octupole deformed \cite{Engel2003}. The radioactivity of all
Ra isotopes and the availability of rather small quantities
requires the atoms to be cooled and trapped for experiments to
achieve sufficient accuracy. Cooling and trapping of a few Ra
atoms has been demonstrated recently using the
7s$^2$~$^1$S$_0$~-~7s7p~$^3$P$_1$ intercombination line
\cite{Guest2007}. The stronger 7s$^2$~$^1$S$_0$~-~7s7p~$^1$P$_1$
has a 100 times higher transition strength and therefore offers a
much larger cooling force. Unfortunately for the heavy alkali
earth elements barium (Ba) and Ra the branching of the decay of
the excited $^1$P$_1$ state to the $^1$S$_0$ state respectively
sum of the $^1$D$_2$ and the $^3$D states is 1:330(30) (Ba) and
1:500(80) (Ra). In the lighter alkali earth atoms Ca and Sr the
ratio is 1:100000(1000) and 1:52000(500), and the transition to
the $^1$D$_2$ provides the strongest of the leaks. Therefore
effective repumping of the population trapped in the metastable
D-states to the cooling cycle will be essential for a scheme
involving the strong transition from the ground state. In
addition, repumping needs to go via the same excited $^1P_1$
state, otherwise the atoms will be distributed over even more
metastable states. We observe dark resonances, which connect the
ground state and the metastable states through coherent
interaction of the cooling laser and the repump lasers. These
Raman transitions decrease the efficiency of the cooling process.

Since their first experimental observation in 1976 in sodium
\cite{ALZETTA1976}, coherent dark resonances have been observed
e.g. in alkali atoms between different hyperfine states, in
chromium \cite{BELL1999} and in samarium vapor
\cite{VLADIMIROVA2003}. A review on the developments and status of
the coherent dark state spectroscopy can be found in
references~\cite{ARIMONDO1996,WYNANDS1999}. We have studied such
processes in the heavy alkali earth atom Ba which has a number of
stable isotopes and is accessible with commercial lasers for all
relevant transitions. We expect that the results can be
transferred directly to Ra because of the similar atomic level
scheme.

\begin{figure}
\center
\includegraphics[width = 7.8cm]{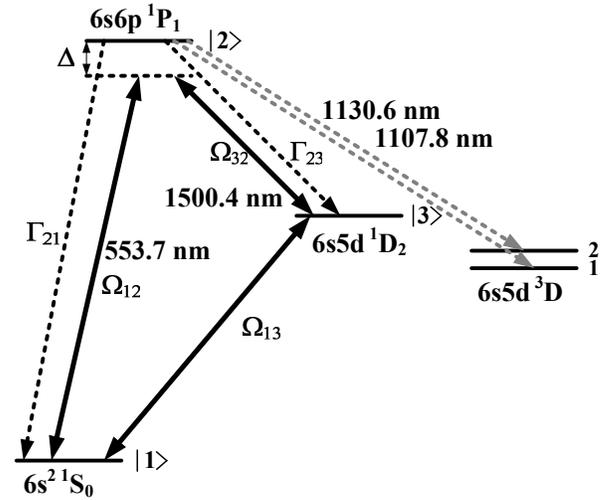}\\
\caption{Ba atomic level scheme for the observation of dark
resonances. The singlet $^1$S$_0$,$^1$P$_1$ and $^1$D$_2$ states
form a ''$\Lambda$ - system'' used here for the dark resonance
measurement. They are referred to as $\mid1\rangle$,
$\mid2\rangle$ and $\mid3\rangle$. The interaction with the
coherent light fields are represented through the Rabi frequencies
$\Omega_{12}$ and $\Omega_{32}$. The decay rates $\Gamma_{21}$ and
$\Gamma_{32}$ are given in Tab. \ref{rabifrequencies}. A common
detuning $\Delta$ for both lasers from the centers of their
resonances is indicated. Other $\Lambda$-systems relevant for
laser cooling with the $^3$D$_{1,2}$ state are also indicated.}
\label{figure1}
\end{figure}

The relevant part of the Ba atom energy level scheme is displayed
in Fig.~\ref{figure1}. The 6s6p~$^{1}$P$_{1}$ excited state decays
to the 6s$^{2}$~$^{1}$S$_{0}$ ground state and weakly to the
6s5d~$^{1}$D$_{2}$ metastable state. If the atom is exposed to a
bichromatic laser field with copropagating green and infrared
laser beams  at $\lambda_1$=553.7~nm and $\lambda_2$=1500.4~nm,
the scattered light contains a strong component $\lambda_1$ and a
small component at $\lambda_2$.

\begin{figure*}[t]
\includegraphics[width = 6.0 cm ,angle = 270]{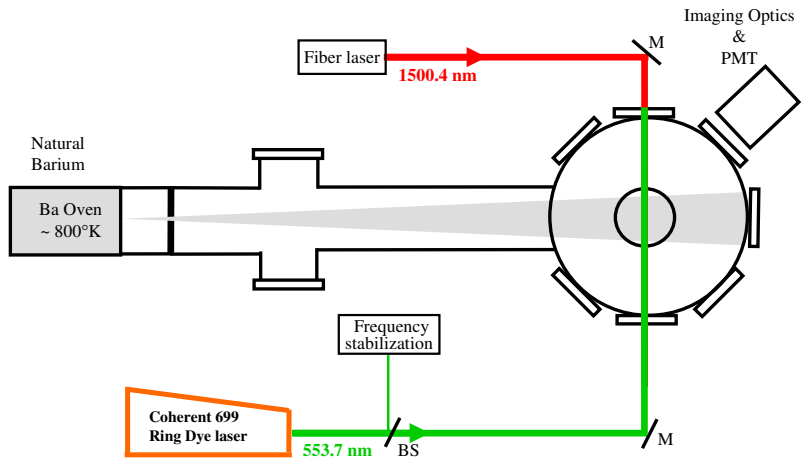}
\caption{A beam of atomic Ba with natural isotope composition
emerges from an oven heated to $\approx$ 800~K. The two
counterpropagating laser beams cross the atomic beam at right
angle 0.6~m downstream, limiting the divergence of the atomic beam
to 30 mrad. Fluorescence from the 6s6p~$^{1}$P$_{1}$ state is
detected with a photomultiplier tube (PMT).} \label{figure2}
\end{figure*}


The experimental setup used to observe the coherent dark
resonances in $\Lambda$-system is depicted in Fig.~\ref{figure2}.
A Ba atomic beam emerges from an oven at T$_{Ba}$=800~K. The
interaction region between the atoms and the two lasers is located
60~cm downstream of the oven. The laser for the excitation of the
6s$^{2}$~$^{1}$S$_{0}$~-~6s6p~$^{1}$P$_{1}$ transition at
$\lambda_1$=553.7~nm is a commercial Coherent 699-21 ring dye
laser with Pyrromethene567 dye pumped by a Nd:YaG laser. Its
linewidth is below 1~MHz. A commercial fiber laser (KOHERAS
Adjustik) at $\lambda_2$ = 1500.4~nm and a linewidth of 50~kHz
excites the 6s5d~$^{1}$D$_{2}$~-~6s6p~$^{1}$P$_{1}$ transition.
The two laser beams are orthogonal to the atomic beam and
counter-propagated each other. They are collimated and are
overlapped along the path of interaction with the atomic beam with
a an angle less than 5~mrad. The beam radius of the laser at
$\lambda_2$ is 1.0(0.2)~mm with a Gaussian beam profile and the
laser beam at $\lambda_1$ is shaped by a rectangular aperture of
size 3$\times$3~mm$^{2}$ to have uniform intensity distribution.
The fluorescence signal from the
6s6p~$^{1}$P$_{1}$~-~6s$^{2}$~$^{1}$S$_{0}$ transition is detected
by a photomultiplier tube with a 550~nm interference filter of
10~nm width in the axis perpendicular to both the atomic and the
laser beams. The fluorescence signal is proportional to the
excited $^1$P$_1$ state population.


The atom-laser interaction in this experiment is modelled as an
atomic three-level $\Lambda$-system,  which coherently interacts
with two light fields. The formalism applied here is similar to
the semi-classical scheme reported to describe the Ba ions
\cite{SHUBERT1995,RAAB2001}. The atom is treated quantum
mechanically in the Heisenberg picture and characterized by the 6
parameters $\mid$i$\rangle$$\langle$j$\mid$ (i,j=1,2 and 3). The
coherent laser fields are treated as classical electromagnetic
waves $\vec{E}_{g} \sin(\omega_{g}t)$ for $\lambda_1$ and
$\vec{E}_{r}~\sin(\omega_{r}t)$ for $\lambda_2$. The two
transitions have a linewidth $\Gamma_{g}$ and $\Gamma_{r}$
respectively. Spontaneous emission of photons is treated as decays
from $\mid$2$\rangle$ to $\mid1\rangle$ and $\mid3\rangle$ with
rates $\Gamma_{21}$ and $\Gamma_{23}$.

The equations of motion for the density matrix $\rho$ of the atom
(Bloch equations) contain terms that model the coherent coupling
with the external light fields by effective Rabi frequencies
$\Omega_{12}$ and $\Omega_{32}$, which denote the strength of the coupling
between the atom and the electric field of the lasers.
\begin{equation}
\Omega_{ij} = \frac{-e E_0}{\hbar}<j|r|i> \label{rabifreq}
\end{equation}
where $e$ is the charge of the electron, $E_0$ is the electric
field of the electromagnetic wave, $\hbar$ is Planck's constant
and the last term is the transition dipole matrix element between
states j and i. The rotating wave approximation, which neglects
the fast oscillating terms, is applied \cite{LOUDON1983}. The
optical Bloch equations where we use the abbreviation
$\tilde{\rho}_{ij}=\rho_{ij}\,e^{i\,\Delta_{ij}\,t}$ is

\begin{eqnarray}
\label{Bloch_eq}
\frac{d\rho_{11}}{dt}& = &~\Gamma_{21}\,\rho_{22}-\frac{i}{2}\,\Omega_{21}^*\,\tilde{\rho}_{12}+ c.c. \nonumber \\
\frac{d\rho_{22}}{dt}& = &-(\Gamma_{21}+\Gamma_{23})\,\rho_{22}+\frac{i}{2}\,\Omega_{12}^*\tilde{\rho}_{21}-\frac{i}{2}\,\Omega_{32}^*\tilde{\rho}_{32}+ c.c.\nonumber \\
\frac{d\rho_{33}}{dt}& = &~\Gamma_{23}\,\rho_{22} + \frac{i}{2}\,\Omega_{32}^*\tilde{\rho}_{23} + c.c.  \nonumber \\
\frac{d\tilde{\rho}_{12}}{dt}& = &-[\frac{\Gamma_{12}+\Gamma_{23}}{2}-i\,\Delta_12] \tilde{\rho}_{12}\\
& & +\frac{i\,\Omega_{12}^*}{2}(\rho_{11}-\rho_{22})+\frac{i}{2}\,\Omega_32\,\tilde{\rho}_{31} \nonumber \\
\frac{d\tilde{\rho}_{13}}{dt}& = &- i [(\Delta_{12}-\Delta{32})\tilde{\rho}_{13}-\frac{\Omega_{32}}{2}\rho_{23}+\frac{\Omega_{12}}{2}\tilde{\rho}_{21}] \nonumber \\
\frac{d\tilde{\rho}_{23}}{dt}& = &
-[\frac{\Gamma_12+\Gamma_{23}}{2}-i\,\Delta_2]
\tilde{\rho}_{23}+\frac{i}{2}\Omega_{32}(\rho_{22}-\rho_{33}) \nonumber \\
& & -\frac{i}{2}\Omega_{12}\tilde{\rho}_{31}
+\frac{i\,\Omega_{32}}{2}(\rho_{33}-\rho_{22}). \nonumber
\end{eqnarray}
The equations are solved numerically for our experimental
conditions. In additions we take the moving of the atoms through
the laser fields into account by considering the finite time the
atoms spent in the respective laser beams. A Maxwell-Boltzmann
velocity distribution at temperature T$_{Ba}$ for the atoms in an
atomic beam is assumed.

Depending on the detuning $\Delta_32$ of the laser at $\lambda_2$
from the 6s5d~$^{1}$D$_{2}$~-~6s6p~$^{1}$P$_{1}$ transition and
the detuning $\Delta_12$ of the laser at $\lambda_1$  from the
6s$^{2}$~$^{1}$S$_{0}$~-~6s6p~$^{1}$P$_{1}$ transition  the atoms
are driven coherently between the $^1$S$_0$ and the
$^1$D$_2$-states without passing through the $^1$P$_1$ state. For
the resonance condition $\Delta=\Delta_12=\Delta_32$ and the
detuning $\Delta$ much larger than the Rabi frequencies
$\Omega_{12}$ and $\Omega_{32}$ the effective Rabi frequency for
the Raman transition can be written as

\begin{equation}
\Omega_{13} = \frac{\Omega_{12}\; \Omega_{32}}{2\Delta}~.
\label{eff_rabi}
\end{equation}

For an intensity of I = 1~mW/cm$^{2}$ of both lasers the estimated
Rabi frequencies (Eq.\ref{rabifreq}) are given in
Table~\ref{rabifrequencies} together with the spontaneous
transition coefficients. However, for effective cooling
$\Delta_{12,32}$ are both on the order of the linewidth of the
transitions and Eq. \ref{eff_rabi} is not applicable. Thus the
full optical Bloch equations need to be evaluated.

\begin{table}[b]
\center
\caption{Wavelengths $\lambda_{ik}$, transition probabilities A$_{ik}$
\cite{NIGGLI1987,BIZZARI1990} and Rabi frequencies $\Omega_{ik}$
at a laser beam intensity I = 1~mW/cm$^2$ for the relevant transitions in Ba. }
\begin{tabular}{|c|c|c|c|}
\hline
\multicolumn{1}{|c|}{Transition}& \multicolumn{1}{|c|}{$\lambda_{ik}$ }&
\multicolumn{1}{|c|}{$\Gamma_{ik}$ }& \multicolumn{1}{|c|}{$\Omega_{ik}$}\\
 \multicolumn{1}{|c|}{}& \multicolumn{1}{|c|}{[nm]}&\multicolumn{1}{|c|}{[10$^{8}$ s$^{-1}$]}& \multicolumn{1}{|c|}{[rad/s]}\\
   \hline\hline
 6s$^{2}$~$^1$S$_0$~-~6s6p~$^1$P$_1$ &  553.7 & 1.19(1) & $22\cdot 10^6$ \\
   \hline
 6s5d~$^1$D$_2$~-~6s6p~$^1$P$_1$ & 1500.4 & 0.0025(2)& $4.7\cdot 10^6$ \\
   \hline
 6s5d~$^3$D$_2$~-~6s6p~$^1$P$_1$ & 1130.6 & 0.0011(2)& $2.0\cdot 10^6$ \\
   \hline
 \hspace{1mm}6s5d~$^3$D$_1$~-~6s6p~$^1$P$_1$ \hspace{1mm}& \hspace{1mm}1107.8 \hspace{1mm} & \hspace{1mm} 0.000031(5) \hspace{1mm} & \hspace{1mm} $0.32\cdot 10^6$ \hspace{1mm} \\
   \hline
\end{tabular}
\label{rabifrequencies}
\end{table}

\begin{figure}[tb]
\center
\includegraphics[width = 8.0cm]{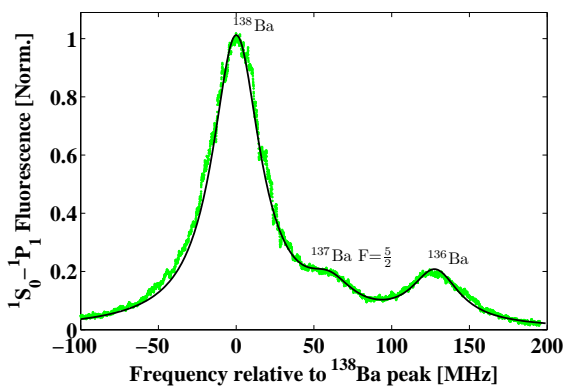}\\
\caption{Green fluorescence spectrum from the $^{1}$P$_{1}$ state
detected with a beam of natural Ba in the absence of the laser at
$\lambda_2$. The experimental spectrum is well described by a
numerical solution of the optical Bloch equations (solid line)
when the spectra are normalized to the peak amplitude.}
\label{figure3}
\end{figure}

\begin{figure}[h]
\center
\includegraphics[width= 8.0cm]{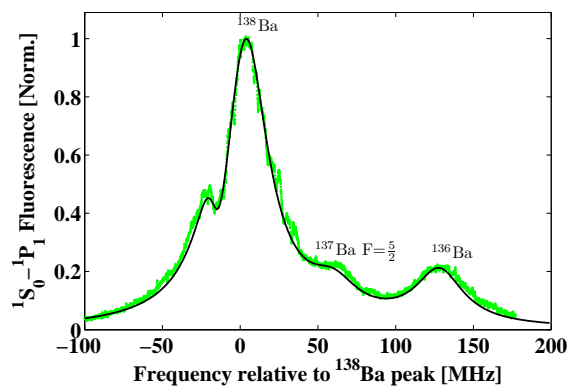}\\
\caption{Green fluorescence spectrum from the $^{1}P_{1}$ state
detected with in beam of natural Ba. A numerical solution of the
optical Bloch equation using $\Omega_{12}$~=~120~$\cdot 10^6$
rad/s and $\Omega_{32}$~=~100~$\cdot 10^6$ rad/s describes the
experimental data well. \label{figure4}}
\end{figure}

The fluorescence spectrum observed shown in Fig.~\ref{figure3} is
obtained by scanning the frequency of the laser at $\lambda_1$ in
the absence of the laser at $\lambda_2$. Contributions from
different isotopes of Ba are visible. The solid line in the figure
represents the numerical solution of Eqns. (\ref{Bloch_eq}). The
spectra are normalized to the peak amplitude of the $^{138}$Ba
resonance. The other isotope contributions are according to their
natural abundance and their known isotope shift
\cite{ba_isoshift}.

Fig. \ref{figure4} gives the fluorescence spectrum obtained with
the laser at $\lambda_2$ (red) detuned by
$\Delta_2$~=~-13.5(5)~MHz with respect to the center of the
6s5d~$^{1}D_{2}$~-~6s6p~$^{1}P_{1}$ transition in the $^{138}$Ba
isotope. The frequency of the laser at $\lambda_1$ is scanned
across the resonance. The laser power at $\lambda_2$ is 10(1)~mW
yielding an intensity of 320(120)~mW/cm$^{2}$. This corresponds to
a Rabi frequency $\Omega_{32}$~=~80(30)~$\cdot 10^6$~rad/s. The
laser at $\lambda_1$ at a power of 2.2(1)~mW, or an intensity of
24(1)~mW/cm$^{2}$, we yield a Rabi frequency of
$\Omega_{12}$~=~120(3)~$\cdot 10^6$~rad/s. The laser linewidth for
the two lasers is much less than the Rabi frequencies and the
decay rate of the 6s6p~$^{1}$P$_{1}$ state. In the calculations a
laser linewidth of $\Gamma_g$~=~1~MHz at $\lambda_1$ and
$\Gamma_r$~=~50~kHz at $\lambda_2$ is assumed. The spectra are
normalized to the peak amplitude. The only adjustable parameter is
the ratio of the coherent Raman transition fraction to the regular
fluorescence spectrum due to atoms not passing through both of the
laser beams. The latter contributes 45(15) $\%$ to the $^{138}$Ba
fluorescence.

The dip on the left side of the resonance in $^{138}$Ba is the
two-photon Raman transition. The fluorescence is reduced because
of the direct coupling of the 6s$^{2}$~$^{1}$S$_{0}$ ground state
and the 6s5d~$^{1}$D$_{2}$ state. The solid lie is the numerical
solution with the parameter for detuning and the Rabi frequency
given above. This demonstrates that the numerical approach fully
describes the atomic response, and such a measurement can be used
to extract the Rabi frequencies $\Omega_{12}$ and $\Omega_{32}$.

\begin{figure}[tb]
\center
\includegraphics[width = 8.0cm]{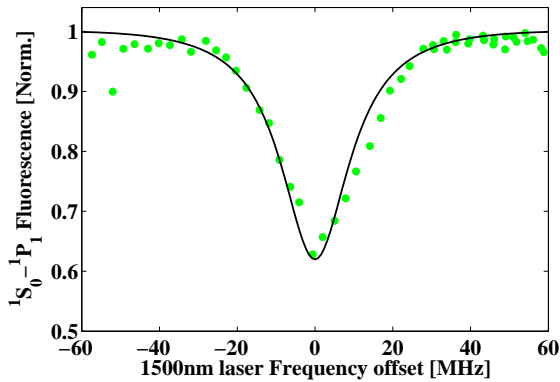}\\
\caption{The measured dark resonance spectrum for the
$^{1}$P$_{1}$ state. Here the laser at $\lambda_1 =$553.7~nm  is
frequency locked to resonance peak in $^{138}$Ba and the laser at
$\lambda_2 =$1500.4~nm laser is scanned.} \label{figure5}
\end{figure}

We have observed the dark resonance in a second measurement. Here,
the green laser at $\lambda_1$ was frequency locked to the peak of
the transition in the $^{138}$Ba isotope using a Doppler shift
method~\cite{DAMMALAPATI2006} and the infrared laser at
$\lambda_2$ was scanned across the resonance (Fig.~\ref{figure5}).
As long as the infrared laser detuning $\Delta_{32}$ is much
larger than the linewidth we observe unperturbed scattering from
the 6s$^{2}$~$^{1}$S$_{0}$~-~6s6p~$^{1}$P$_{1}$ transition. When
the detuning $\Delta_{32}$ approaches the resonance the scattering
rate decreases, because of coherent population transfer to the
metastable 6s5d~$^{1}$D$_{2}$ state. The width of the observed dip
is 20(1)~MHz, in good agreement with the linewidth from the
numerical calculation of 19~MHz for this transition. This
agreement indicates that residual Doppler broadening due to
improper overlap of the laser beams was small. The solid, smooth
line represents the numerical solution of the optical Bloch
equations with $\Omega_{12}$=100~$\cdot 10^6$rad/s,
$\Delta_12$=0~MHz and $\Gamma_1$=$1$~MHz for the laser at
$\lambda_1$; $\Omega_{32}$=120~$\cdot 10^6$rad/s, and
$\Gamma_2$=$50$~kHz for the laser at $\lambda_2$. The fluorescence
is normalized to the peak values and again get a good agreement
between the experiment and our model.

Thus we can expect that Rabi frequencies of several $10^7$~rad/s
can be achieved with a laser beam diameter of 5 mm and the
available laser power of 10~mW from the fiber laser. To achieve
the same Rabi frequency for the
6s5d~$^3$D$_2$~-~6s6p~$^{1}$P$_{1}$ transition the laser intensity
has to be about 5 times larger and for
$^3$D$_1$~-~6s6p~$^{1}$P$_{1}$ transition more than 100 times
larger (Table~\ref{rabifrequencies}).

We have reported first measurements on dark resonances in atomic
Ba. The observations in a three level subsystem can be well
described with a model using numerical solutions of the optical
Bloch equations. The accuracy of the determination is sufficient
for the evaluation of the feasibility of laser cooling of barium.
For a laser cooling scheme for Ba the repumping from $^3$D states
have to be included. They will also exhibit coherent Raman
transitions. The full 5-level system can be modelled with the same
approach and solved numerically. In conclusion we expect that
using the strong transition at $\lambda_1$ and the repumping
through the same excited state can lead to a more effective
cooling of heavy alkali earth atoms like Ba and Ra than the
demonstrated laser cooling on the 7s$^2$~$^1$S$_0$-7s7p~$^3$P$_1$
intercombination line in Ra \cite{Guest2007}. This will allow for
more stringent limits on permanent electric dipole moments in
atomic systems.

This work was supported by the \textit{Stichting voor Fundamenteel
Onderzoek der Materie (FOM)} under programme 48 (TRI$\mu$P). One
of us (U.D.) wishes to ackowledge the support through a
\textit{Ubbo Emmius} PhD student fellowship.

\end{document}